\documentclass[slac_one]{revtex4}
\usepackage{graphicx}
\usepackage{fancyhdr}
\newcommand{\GX}{G$\chi^2$}
\newcommand{\del}{\ensuremath{\partial}}
\newcommand{\mum}{\ensuremath{\mu\rm{m}}}
\begin{document}

\title{{\small{Hadron Collider Physics Symposium (HCP2008),
Galena, Illinois, USA}}\\ 
\vspace{12pt}
The Alignment of the ATLAS Silicon Tracker} 

%

\author{O.~Brandt on behalf of the ATLAS collaboration}
\affiliation{University of Oxford, Keble Road, Oxford, OX1 3RH, UK}

\begin{abstract}
The alignment of tracking detectors is crucial for the physics programme at the LHC, especially for precision measurements like the $W$ mass. The alignment of the ATLAS silicon tracking detectors with a total of 5832 modules poses an unprecedented challenge. In this poster, two track-based alignment algorithms at ATLAS will be covered: the Global $\chi^2$ Algorithm based on $\chi^2$ minimisation, and the Robust Alignment Algorithm based on centering residual distributions. On a time scale of minutes and hours, the geometry of the silicon tracking detectors can be monitored with the Frequency Scanning Interferometry system. An overview of this system will also be given.
\end{abstract}

\maketitle

\thispagestyle{fancy}

\section{The ATLAS Inner Detector and Its Alignment} 
\begin{figure*}[t]
\centering
\includegraphics[width=135mm]{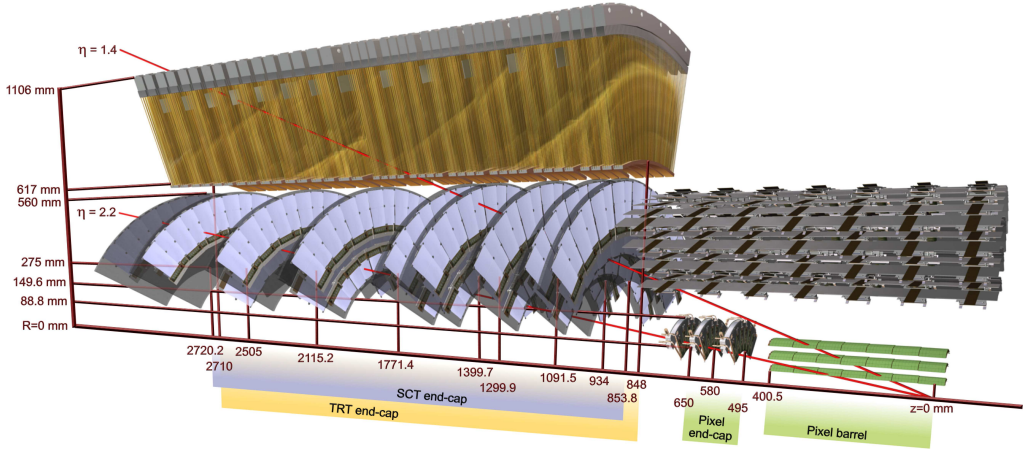}
\vspace{-3mm}
\caption{\label{fig:InDet}
View of one segment of the Inner Detector of ATLAS with two tracks.
}
\vspace*{-4mm}
\end{figure*}

The ATLAS Inner Detector (ID), depicted in Figure~\ref{fig:InDet}, is designed to measure the momentum of charged particles from their bending radius in the 2\,T solenoidal magnetic field on the one hand, and to provide a good pattern recognition for efficient vertexing on the other hand. 
It consists of the following subdetectors~\cite{bib:tdr, bib:IDtdr, bib:detectorPaper}:
\begin{itemize}
\vspace*{-2mm}
\item The Pixel detector is a silicon tracking device with a uniform pixel size of 50$\,\mu$m $\times$ 400$\,\mu$m. It provides a 2-dimensional measurement with a resolution of 14$\,\mu$m $\times$ 115$\,\mu$m for single-pixel hits~\cite{bib:tdr}. This can be further improved with hit clustering, especially for low incidence angles~\cite{bib:detectorPaper}. The barrel part of the Pixel detector is comprised of three layers of wafers, whereas the endcaps have 3 disks each.
\vspace*{-2mm}
\item The Semi-Conductor Tracker (SCT) is a silicon strip detector with 4 barrel layers and 9 endcap disks. Each barrel module features two single-sided back-to-back strip wafers with 80$\,\mu$m pitch, affixed at 40\,mrad stereo angle. At normal track incidence angle, this setup provides for a space point resolution of approximately 17\,\mum\ perpendicular to the strips, and 580\,\mum\ along them~\cite{bib:tdr,bib:detectorPaper}. Endcap modules have a similar setup but a fan-out strip structure~\cite{bib:tdr}.\vspace*{-2mm}
\item The Transition Radiation Tracker is a gaseous straw tube detector. It consists of approx. 300\,000 individual straw tubes with a 4\,mm diameter, providing a per-tube resolution of approximately 130$\,\mu$m  perpendicular to the anode wire~\cite{bib:detectorPaper}. The straws of the TRT are arranged into 96 barrel and 2 $\times$ 28 endcap modules.
\vspace*{-2mm}
\end{itemize}

To fully profit from the performance of the ATLAS ID, the best possible knowledge of module positions is essential. After assembly and optical survey, this can be achieved by two means: \textit{track} and \textit{hardware} based alignment techniques, covered in Section~\ref{sec:trackBasedAlignment} and Section~\ref{sec:hardwareBasedAlignment}, respectively. 

Generally, for each of the silicon modules, 6 alignment degrees of freedom (DoF) can be defined: 3 translations and 3 rotations. For the silicon detectors these are the two in-plane measurement directions, local $x$ and local $y$ along the more and less precise measurement directions.
Local $z$ is orthogonal to the first two, and points away from the interaction point.
At the same time, local $x$, $y$, and $z$ are the module rotation axes. 

\section{Track-based Alignment}\label{sec:trackBasedAlignment}
Currently, at ATLAS there are 3 alignment algorithms for the silicon detectors: the Robust Alignment Algorithm (RA) described in Subsection~\ref{ssec:RA}, the Global $\chi^2$ Algorithm (\GX) introduced in Subsection~\ref{ssec:GX2}, and the Local $\chi^2$ Algorithm which shall not be discussed here. The alignment algorithm for the TRT~\cite{bib:trtAlign} will not be covered here either. All the alignment algorithms are part of the official ATLAS software~\cite{bib:athena}. Efforts are underway to consolidate these 3+1 approaches into one single algorithm.

All the above approaches are based on track residuals $r$, which are canonically defined as the distance between the intersect of the track with a given module and the measured hit position. The main postulate of track-based alignment approaches is that for a perfectly aligned detector all the residual distributions will be centered around zero and will have a width determined by multiple scattering and the intrinsic detector resolution only. Accordingly, for the quality of alignment a figure of merit calculated from the residuals can be defined, e.g. the $\chi^2$ for the \GX{} algorithm (see Subsection~\ref{ssec:GX2}).


During the alignment procedure it can happen that a local minimum of the alignment figure of merit is reached while some global distortions of the detector remain, resulting in biased track parameter distributions. Such distortions are called \textit{weak modes}. 
Some examples are ``clocking'' ($\delta\phi=\beta /R$), elliptical distortions in transverse plane, ``telescope'' ($\delta Z\sim R$), and $\phi$-dependent sagitta distortions ($\delta X=\lambda R +\gamma R^2$).
They can be tackled to some extent with infinite track statistics. For some of those deformations, requiring tracks to come from the beamline as well as imposing direct constraints from an independent measurement can be very helpful. 
Cosmic events which traverse both hemispheres of the detector with a continous, off-beamline helix can be very effective for the barrel part of the detectors, whereas beam-gas and beam halo events can be used to eliminate weak modes in the endcaps.


\subsection{Robust Alignment}\label{ssec:RA}
Following the main postulate of track-based alignment approaches, the main concept of the Robust Alignment Algorithm (RA)~\cite{bib:RA} is based on centering $r_x$, $r_y$ residual distributions for each of the individual modules.
Intuitively, for a perfectly aligned detetector with only a single misaligned module, this will give the desired result. However, for a situation where the whole ID is misaligned, correlations between misaligned modules are propagated by the track fit. They are taken into account via iterations. Typically, $N_{\rm iter}=o(10)$ are necessary. Two types of residuals can be used: \textit{biased}, where all the hits are considered in the track fit, and \textit{unbiased}, where the hit of the given residual is excluded.


In the RA algorithm, the so-called overlap residuals $o$ play a special role. They are defined as the difference between two residuals of the same track in two neighbouring modules in the same layer/disk: $o_{jk}\equiv r_j-r_j$, where $k=x,y$ designates the edge of the module where the overlap occurs, and $j=x,y$ the residual type. Their advantageous property is the small distance the track traverses between the modules, which reduces effects due to multiple scattering. To profit from this, the RA algorithm offers the possibility to assign them an increased weight $w_o>1$ compared to regular residuals.

Overall, the alignment corrections $a$ are derived using the formula:
\begin{equation}
  a_j=-{\sum}_{\rm tracks}\Big{(}r_j + w_o\cdot{\sum}_{k=x,y}o_{jk}\Big{)}\,,\quad j=x,y
\end{equation}

\subsection{Global $\chi^2$ Method}\label{ssec:GX2}
The Global $\chi^2$ Algorithm (\GX)~\cite{bib:GX2} is based on the minimisation of the figure
\begin{equation}
  \chi^2\equiv{\sum}_{\rm tracks}\,r^TV^{-1}r\,,
\end{equation}
with respect to alignment parameters $a$. Here, $r$ is the residual vector for a given track, and $V$ its covariance matrix. Certainly, this quantity depends via the residuals $r$ on the fitted track parameters $\pi$. Linearising this expression around the minimum under the assumption of small alignment corrections, the general solution is given by:
\begin{equation}
  \delta a = -{\Big (}{\sum}_{\rm tracks}\frac{dr^T}{da}V^{-1}\frac{dr}{da}{\Big )}^{-1}{\sum}_{\rm tracks}\frac{dr^T}{da}V^{-1}r\,,
  \quad{\rm with\quad}
  \frac{dr}{da} = \frac{\del r}{\del a} + \frac{\del r}{\del\pi}\frac{d\pi}{da}\,.
\end{equation}
Similarly, $d\pi/da$ can be obtained for one single track:
\begin{equation}
  \delta\pi = -{\Big (}\frac{dr^T}{d\pi}V^{-1}\frac{dr}{d\pi}{\Big )}^{-1}\frac{dr^T}{d\pi}V^{-1}r\,.
\end{equation}
The main advantage of the \GX{} algorithm is the proper treatment of correlations between residuals of the same track in different modules, and the fact that it generically can treat the full $a$ vector, i.e. 6 DoF per module. The technical difficulty of finding the (mathematically) exact solution with the \GX{} algorithm lies with the necessity to solve a system of $N_{\rm DoF}=N_{\rm DoF\,p.\,mod.}\times N_{\rm mod.}$ linear equations, which is inherently singular. Various preconditioning and fast solving techniques like MA27~\cite{bib:ma27} have been implemented, as documented in~\cite{bib:fastSolvers}.

\subsection{On-surface Cosmics Alignment Results}\label{ssec:sr1}

In June 2006 around 400\,k cosmic ray events were recorded by parts of the SCT and TRT detectors in the ATLAS surface building SR1~\cite{bib:sr1}. 468 out ouf 2112 SCT barrel modules arranged in two opposite full-length wedges were read out, as well as 12 out of 96 TRT barrel modules. No pixel detector was present. 
The trigger signal was given by three scintillator plates operated in coincidence. With a lead plate between the lower two a momentum cut-off at approximately 100\,MeV was introduced. No magnetic field was used in this setup. 
This challenge was highly important for commissioning of the ID, the tracking, and the alignment. 

\begin{figure*}
\centering
\includegraphics[width=85mm,clip]{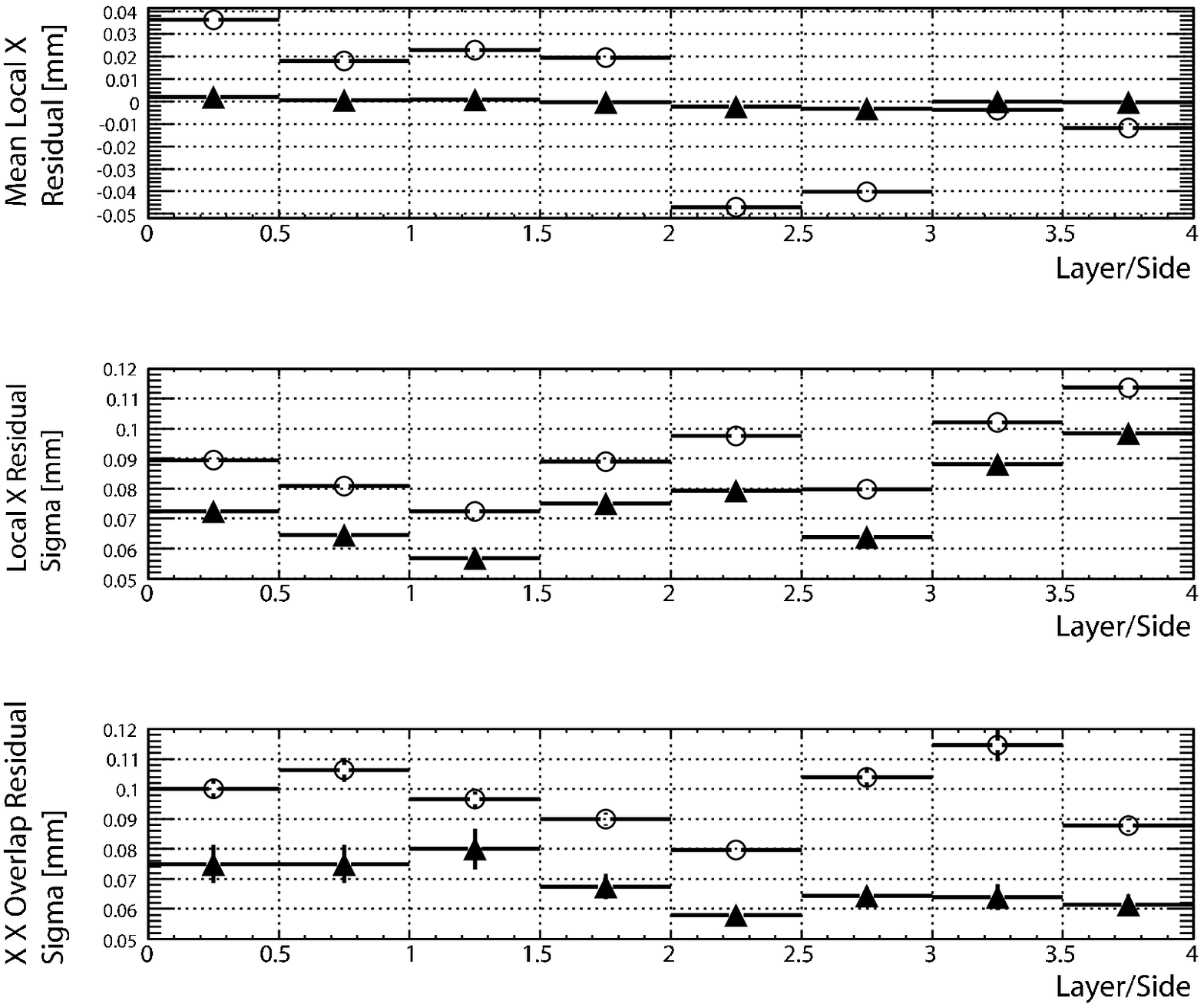}
\includegraphics[width=85mm,clip]{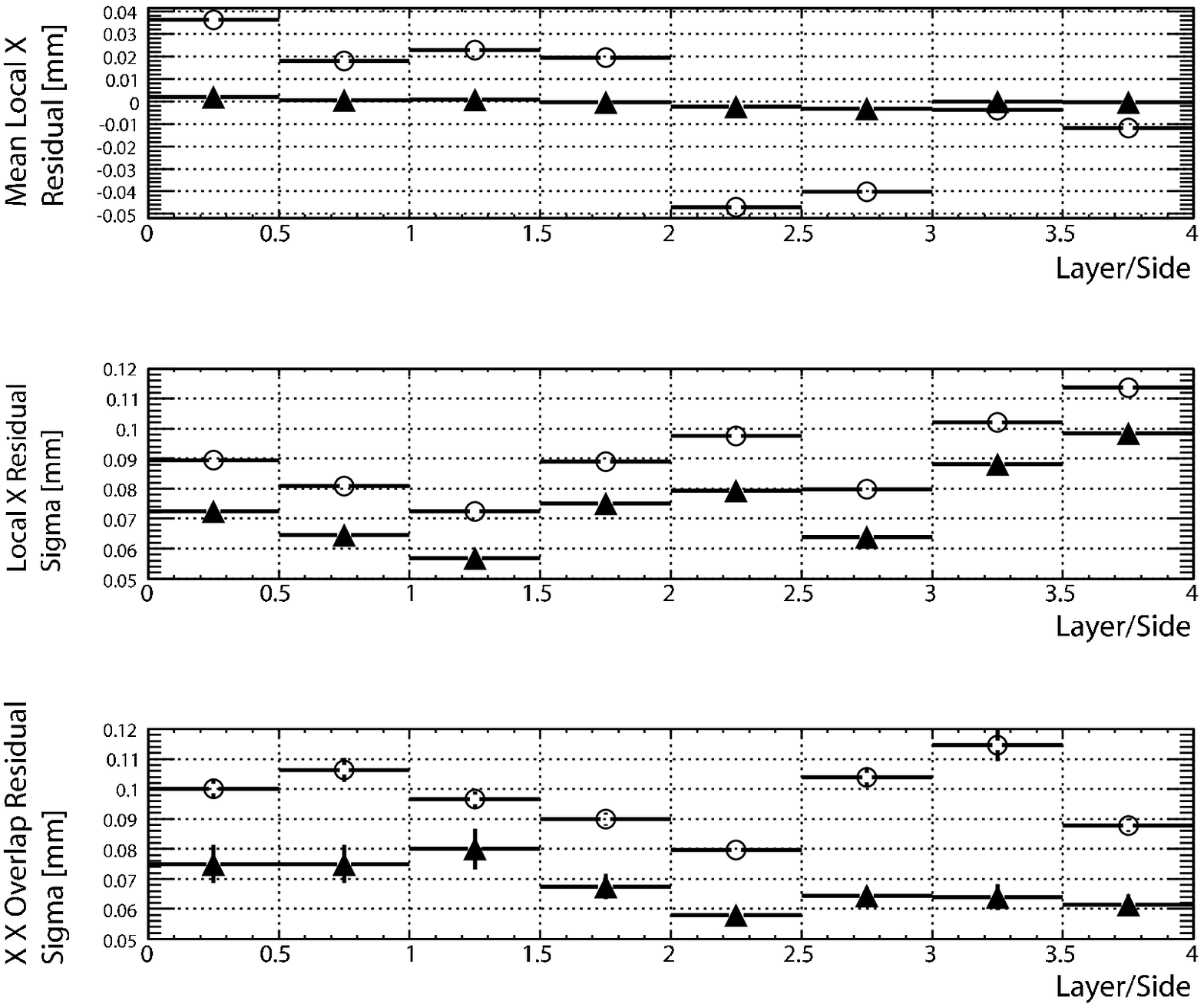}
\vspace{-4mm}
\caption{\label{fig:sr1RA}
Means and $\sigma$-parameters of Gaussian fits to peaks of $r_x$ residual distributions for SR1 cosmic data before (open dots) and after (full triangles) alignment. For each layer $i$ both module sides are shown, labeled as $i$ and $i+0.5$, respectively.
}
\end{figure*}

The RA algorithm was applied on the SR1 cosmics data, and provided fast and reliable results~\cite{bib:RA}. Convergence was reached within 10 iterations. The $r_x$ residual means and resolutions are shown for all four layers of the SCT in  Figure~\ref{fig:sr1RA}. The combined SCT resolution in the peak region improves from 65.1\,$\mu$m to 35.2\,$\mu$m. This compares well to the residual width expected for a perfectly aligned geometry, which is 31.2\,$\mu$m, and is an impressive result given that the RA algorithm aligns only two main DoF for each module.

\begin{figure}
\begin{center}
\begin{picture}(440,110)(0,0)
\put(-5,0){\includegraphics[width=1.4in,height=1.5in]{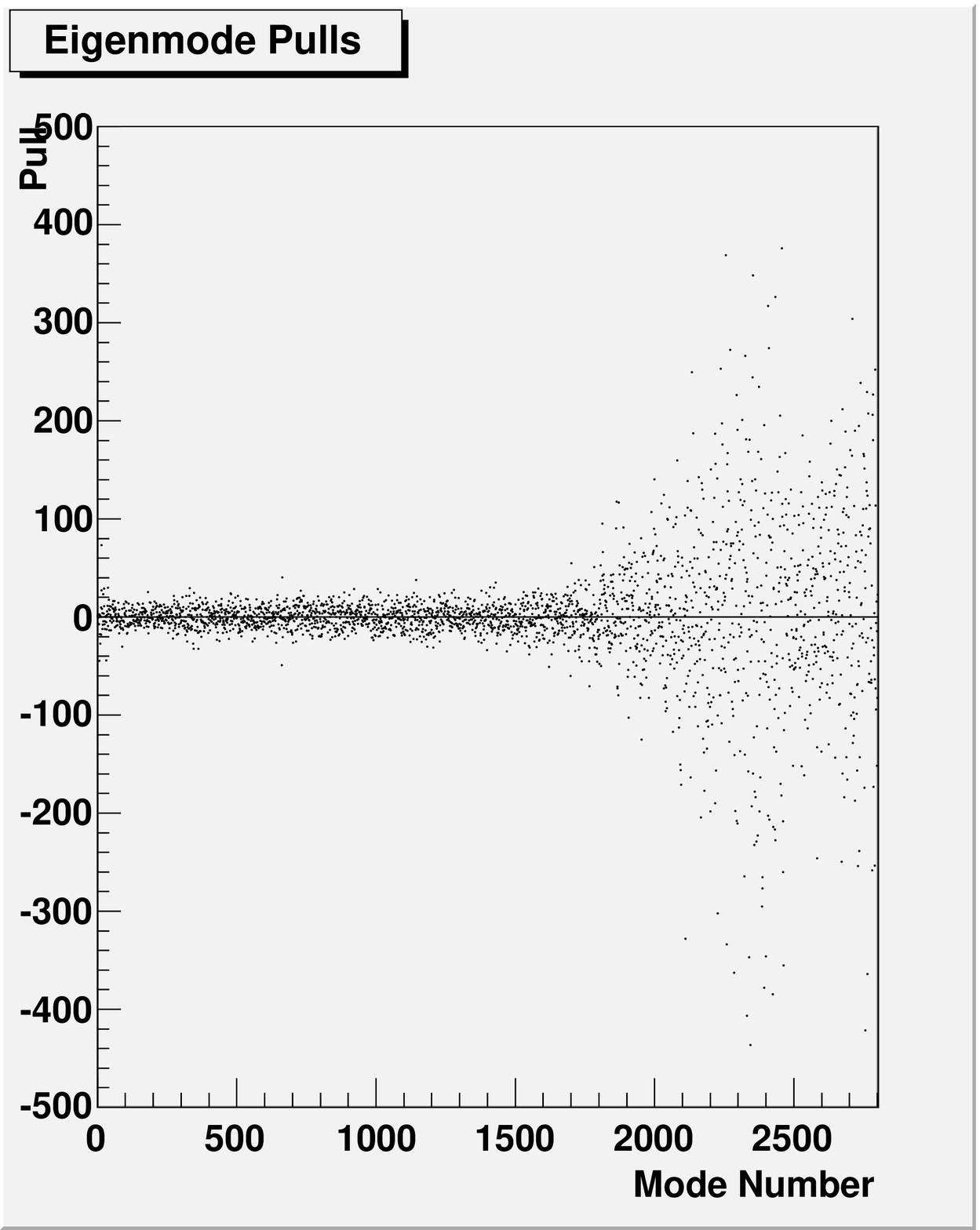}}
\put(110,0){\includegraphics[width=1.5in,height=1.5in]{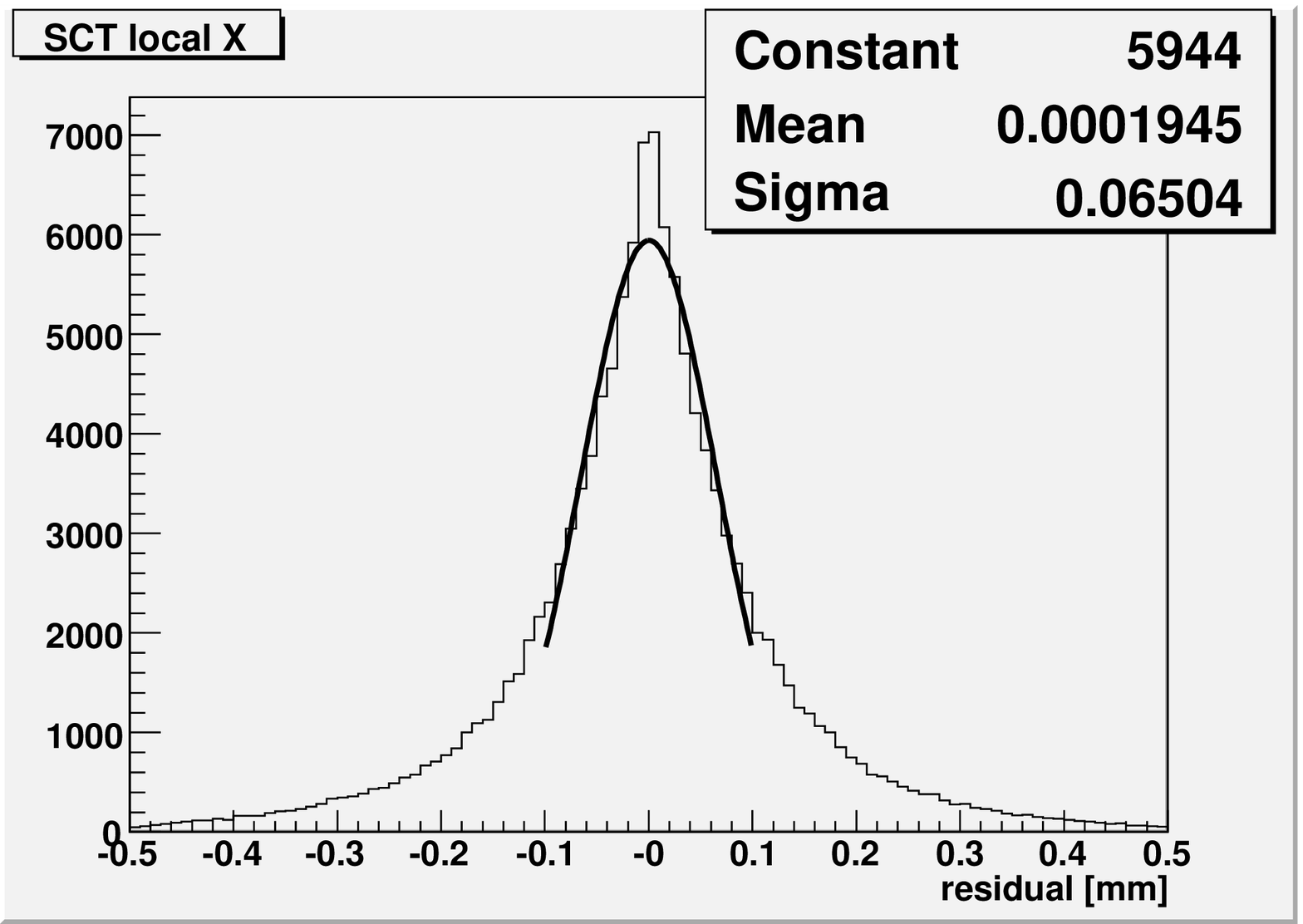}}
\put(245,0){\includegraphics[width=1.4in,height=1.5in]{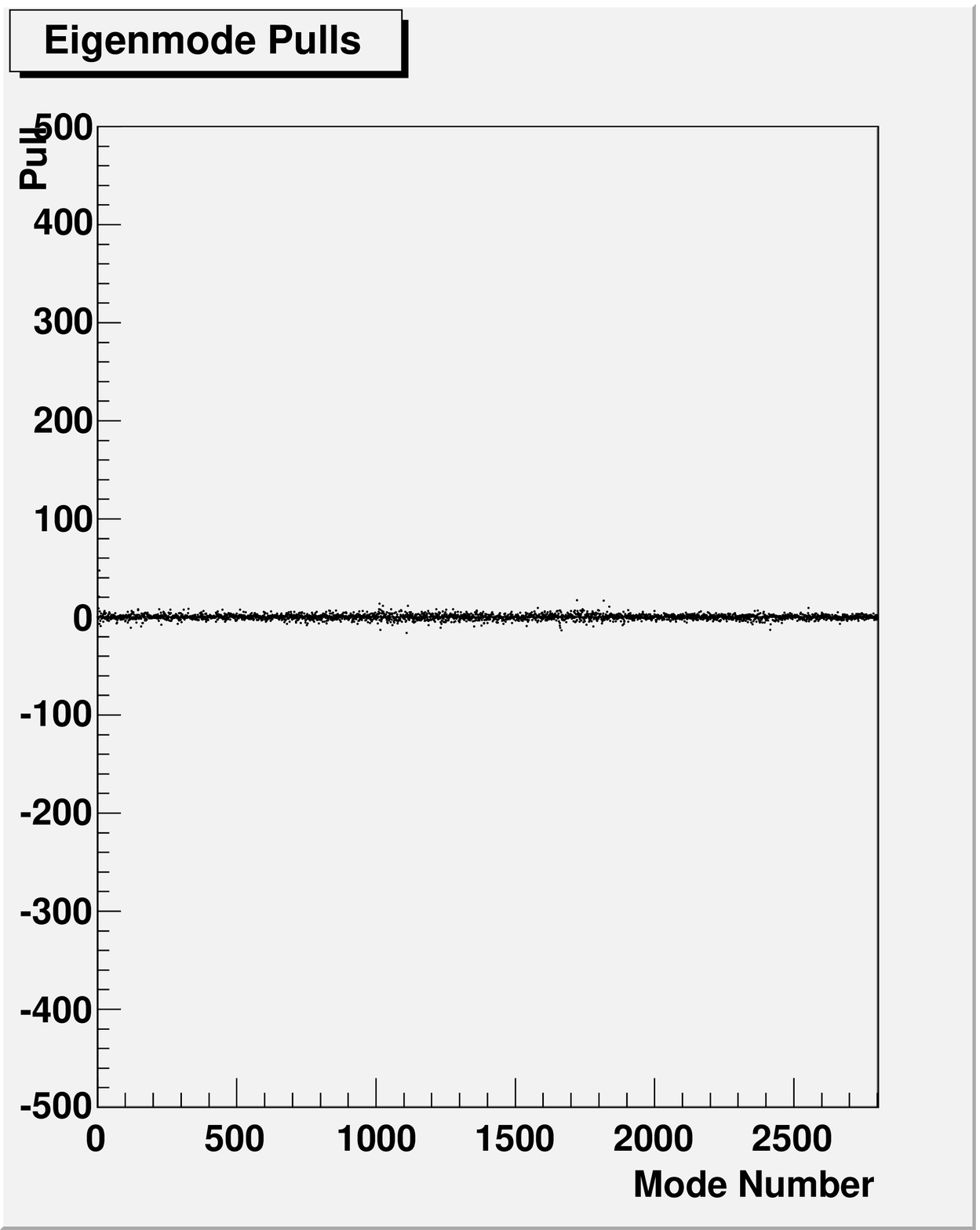}}
\put(360,0){\includegraphics[width=1.5in,height=1.5in]{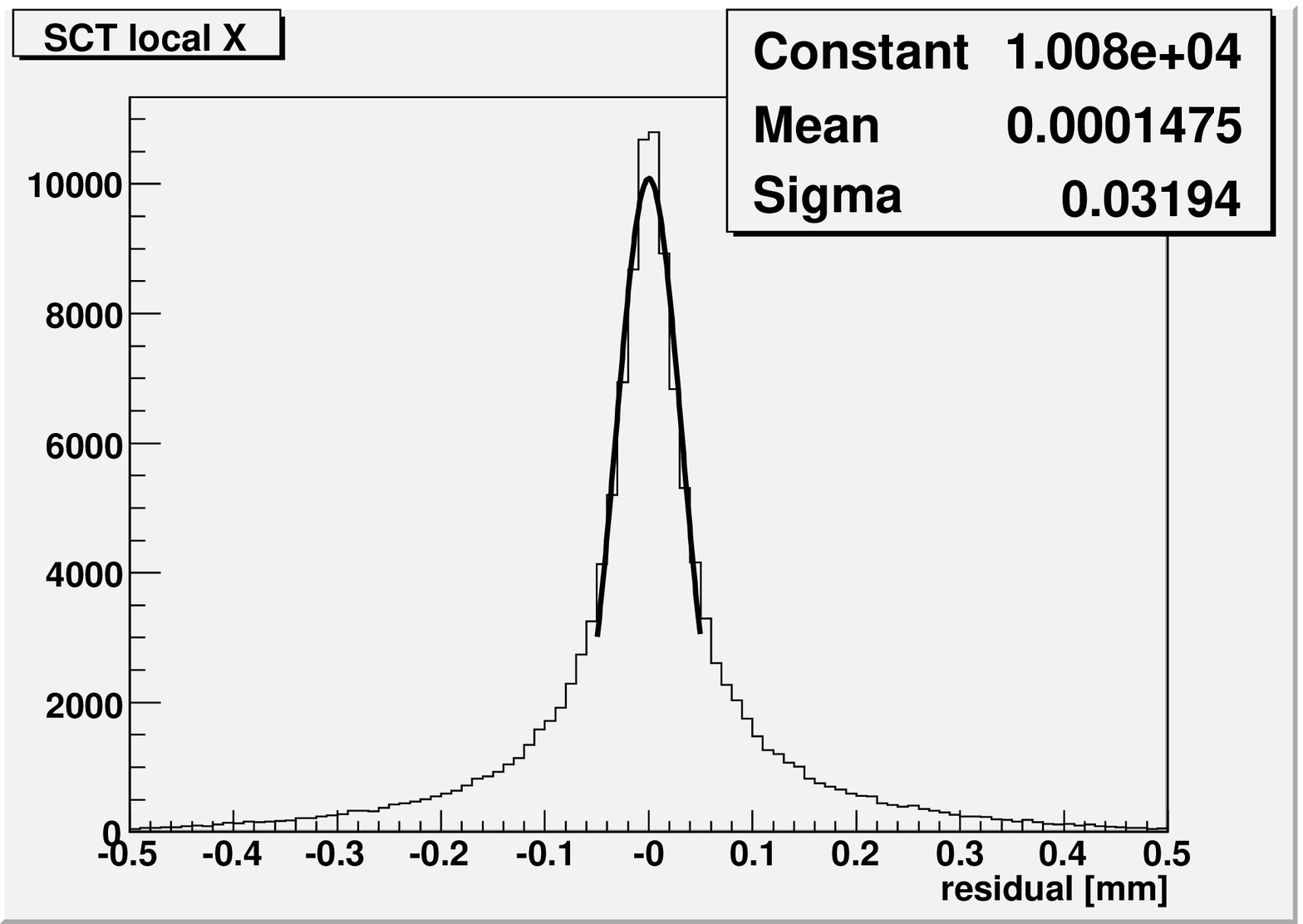}}
\put(97,45){\large $\Leftrightarrow$}
\put(347,45){\large $\Leftrightarrow$}
\put(15,80){($a$)}
\put(130,80){($b$)}
\put(265,80){($c$)}
\put(380,80){($d$)}
\end{picture}
\vspace*{-2mm}
\caption{\label{fig:sr1GX}
Performance of the Global $\chi^2$ algorithm on the SR1 cosmic setup. The left pair of plots shows correction pulls for the 2808 eigenmodes ($a$) and corresponding $r_x$ residual distribution ($b$) before alignment. The right pair of plots ($c$) and ($d$) shows analogous distributions on aligned geometry.
}
\vspace*{-6mm}
\end{center}
\end{figure}

The performance of the \GX{} algorithm was also tested on the SR1 cosmics setup. As this algorithm inherently considers the correlation between modules, convergence was reached after 2 iterations. The results are summarised in Figure~\ref{fig:sr1GX}. For the plots of the correction pulls the diagonal base, defined by the eigenmodes of the detector deformations being orthogonal to each other, was used. The eigenvalues are sorted from left to right by their increasing eigenvalue. The increasing amplitudes of the pulls on the right hand side of plot ($a$) correspond to the misalignment modes that the experimental setup is actually sensitive to. After aligment, the combined width of the residuals is 32\,\mum{}, which is approximately 1\,\mum{} away from the ideal alignment situation.

\subsection{Results with the Full Inner Detector Using Monte Carlo Simulations}\label{ssec:CSC}

The track-based alignment algorithms have been exercised in a full scale scenario with Monte Carlo simulations. For this, the geometry of the detector has been conservatively distorted to the level expected from assembly tolerances. Systematic macroscopic shifts of $o$(mm) of subdetectors were introduced, and the individual modules were randomly misaligned according to a flat distribution with a width of $o$(50\,\mum), depending on subdetector and DoF. Additionally, some weak modes were introduced. More details on the introduced misalignments can be found in~\cite{bib:cscMisal}.

\begin{figure}
\begin{center}
\begin{picture}(440,80)(0,0)
\put(-35,0){\includegraphics[height=30mm,width=83mm,clip]{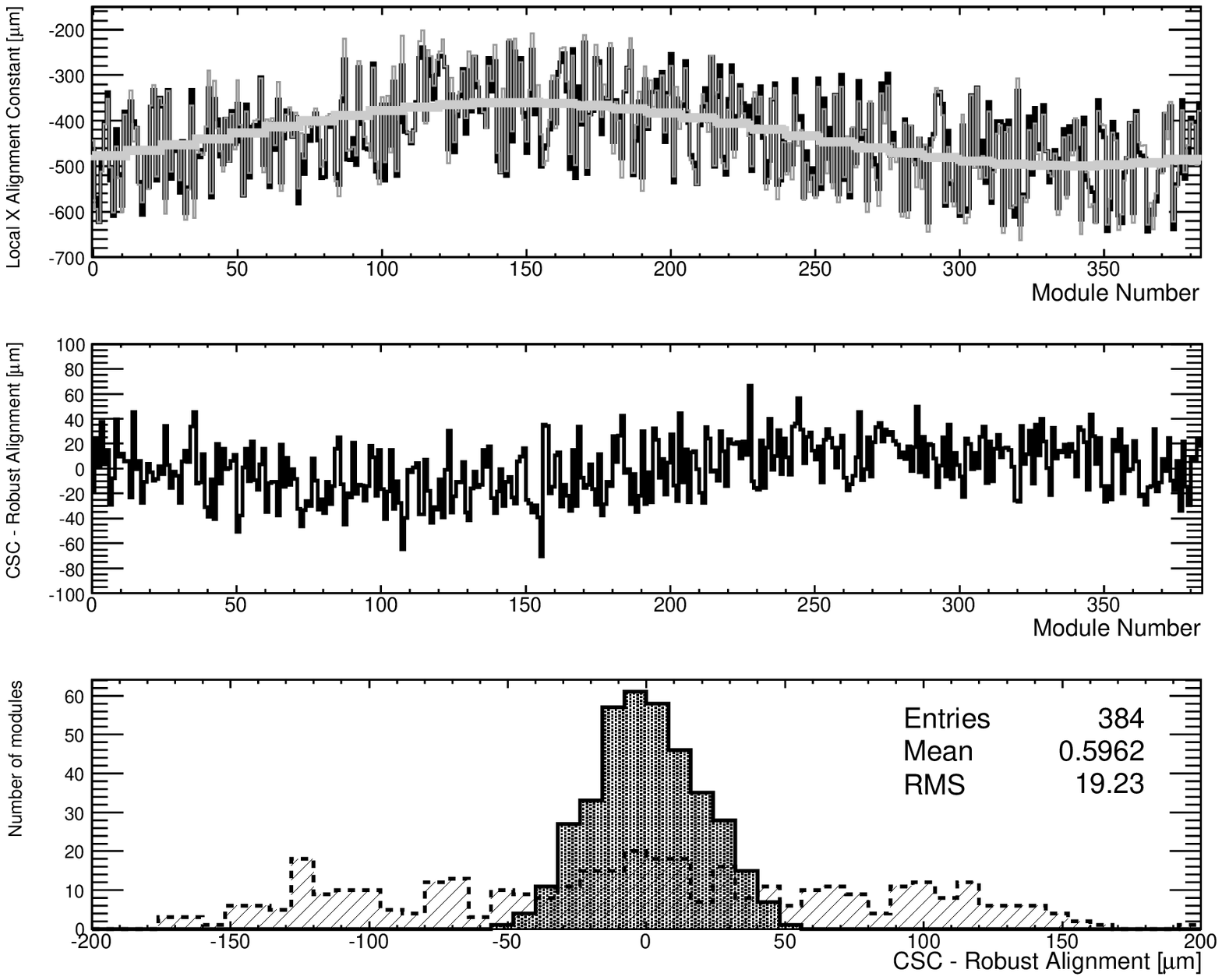}}
\put(205,0){\includegraphics[height=30mm,width=83mm,clip]{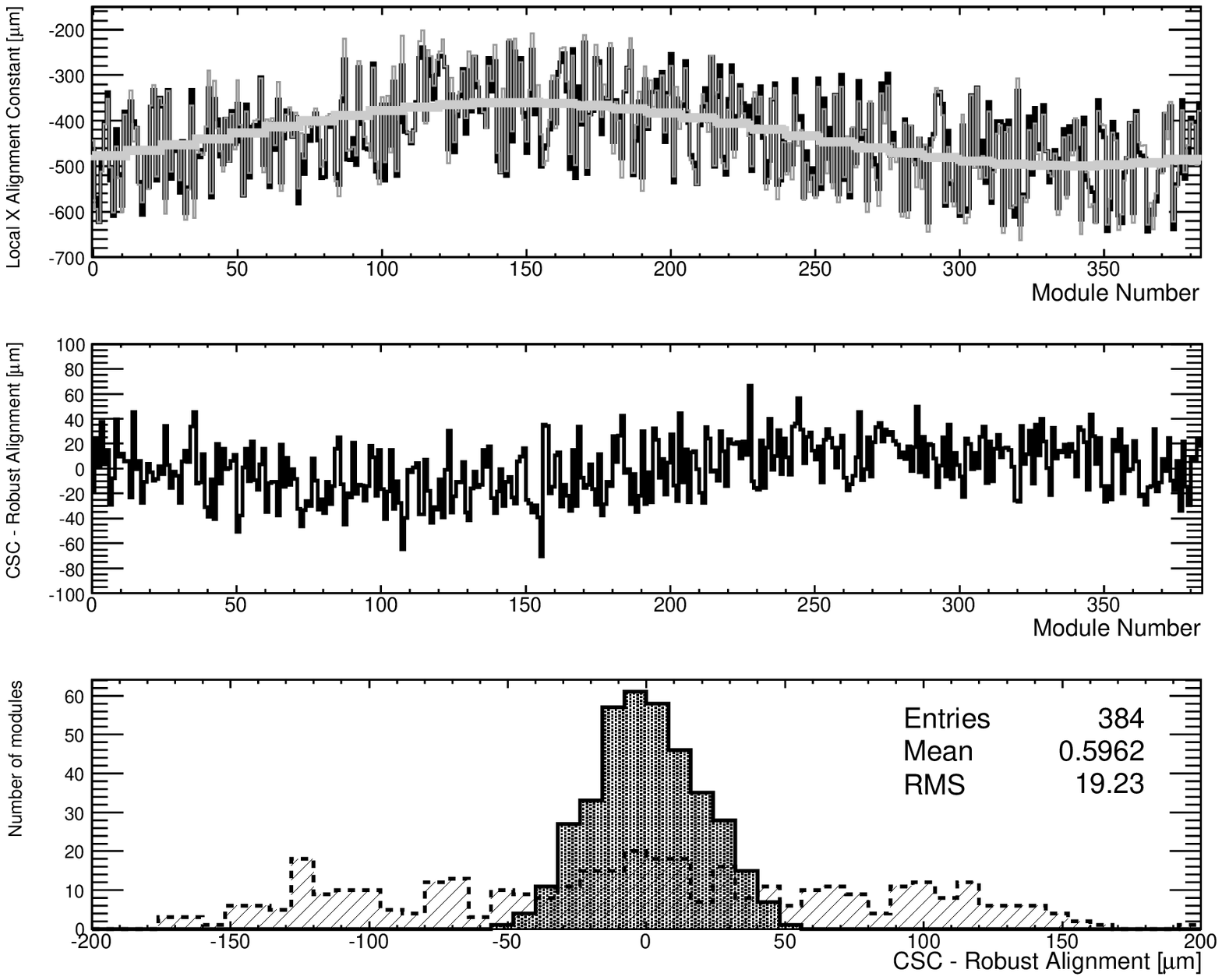}}
\put(-5,70){($a$)}
\put(235,70){($b$)}
\end{picture}
\vspace*{-4mm}
\caption{\label{fig:cscRA}
The performance of the RA algorithm in layer 0 of the SCT barrel for the full ID simulation. In ($a$), the alignment constants derived with RA (fine grey), the true misalignments (black line), and the global misalignments (bold grey) are shown. Figure ($b$) displays the deviations of RA constants from the true misalignments before (light histogram) and after alignment (dark histogram).
}
\end{center}
\end{figure}

The RA algorithm was the first to provide results for this alignment commissioning exercise~\cite{bib:RA}. This could be achieved as this algorithm is naturally scalable and universally applicable. Figure~\ref{fig:cscRA} shows the alignment of layer 0 of the SCT as an example of the performance of the RA. Misalignments of up to $\pm150$\,\mum{} in the local $x$ direction are successfully corrected with a precision of roughly 20\,\mum. These remaining misalignments are due to the limited track statistics, module rotations, as well as global distortions. From a detailed analysis it can be concluded that this quality of alignment should be sufficient for not degrading the intrinsic resolution by more than 20\,\%.

\begin{figure}
\begin{center}
\begin{picture}(440,90)(0,0)
\put(30,0){\includegraphics[height=35mm,width=54mm,clip]{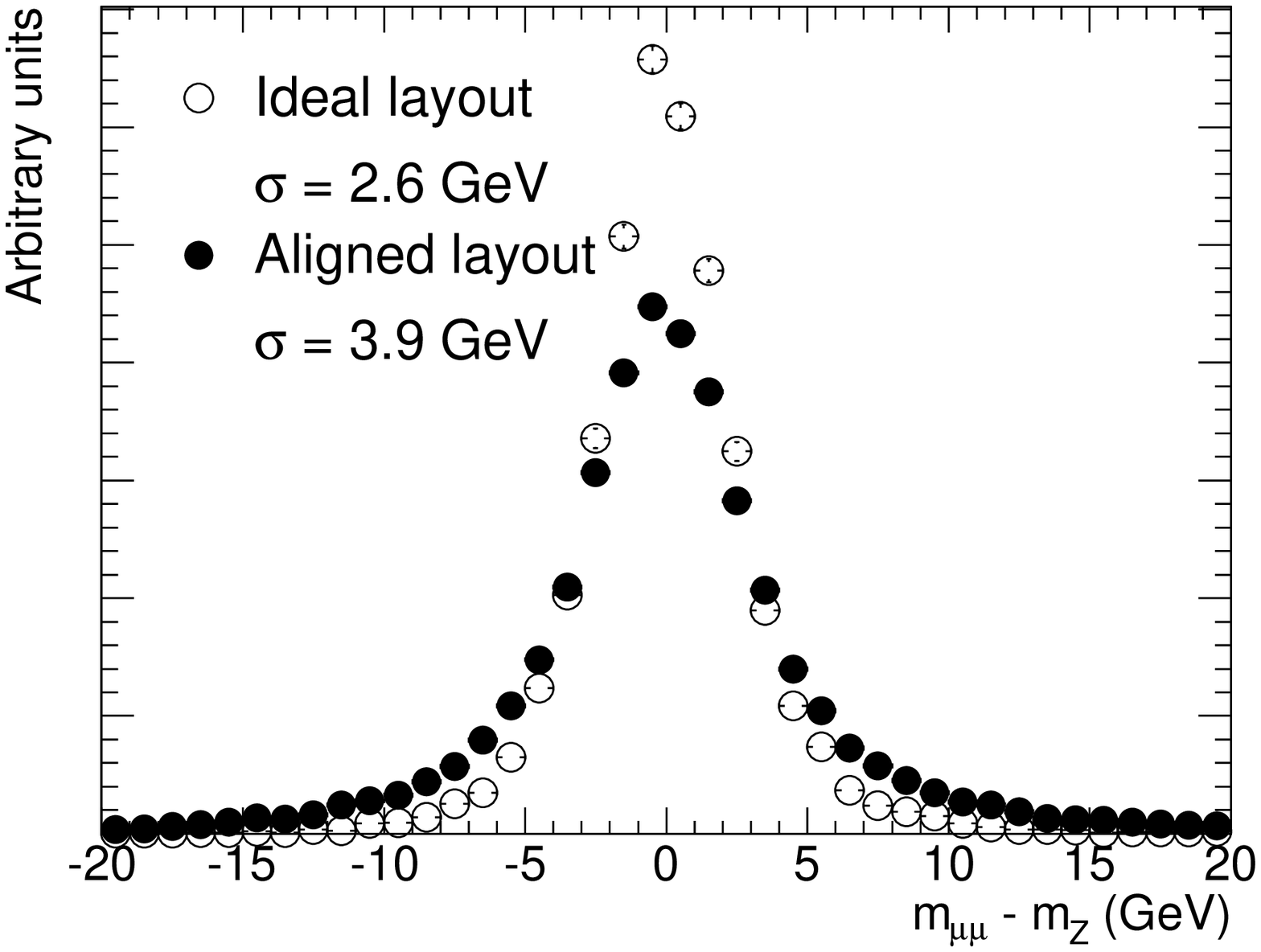}}
\put(235,0){\includegraphics[height=35mm,width=54mm,clip]{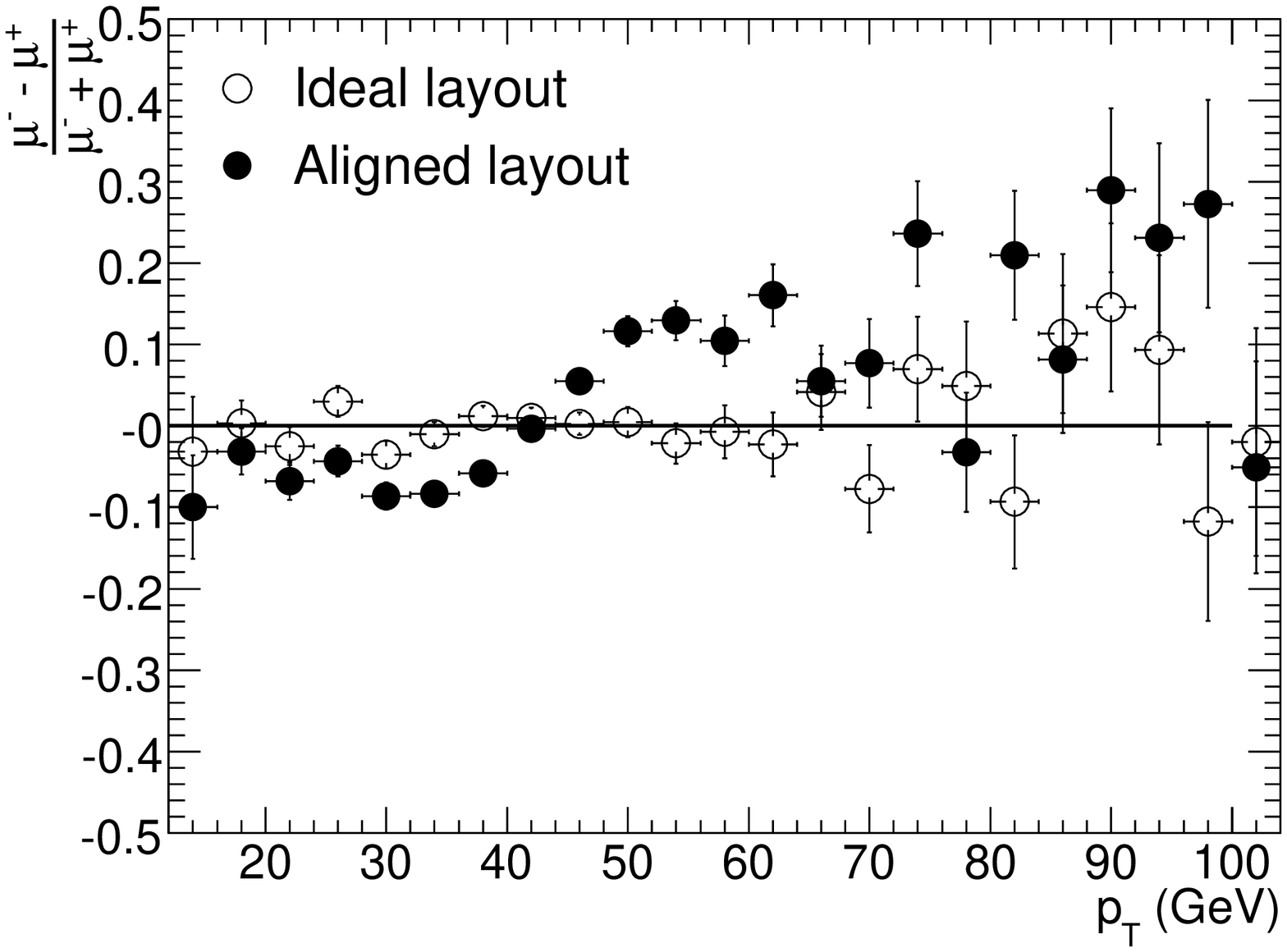}}
\put(55,25){($a$)}
\put(270,25){($b$)}
\end{picture}
\vspace*{-4mm}
\caption{\label{fig:cscGX}
Validation results for the alignment with the \GX{} algorithm. For perfectly known (``ideal'') and the \GX-realigned geometry the difference $m_{\mu\mu}-m_{Z}$ is shown in ($a$), and the charge asymmetry for the same resonance in ($b$).
}
\vspace*{-4mm}
\end{center}
\end{figure}

The simulated ATLAS ID was also aligned using the \GX{} algorithm. For this, tracks have been loosely constrained to the beamline position. The effect of alignment on the $Z\rightarrow\mu\mu$ resonance is shown in Figure~\ref{fig:cscGX} ($a$). While the width for ideal alignment is 2.6\,GeV, the aligned geometry gives 3.9\,GeV. 
This is an impressive result, given that the $Z$ peak was not visible initially.
The charge asymmetry
which is sensitive to ``clocking'' distortions, is shown in bins of reconstructed $p_T^\mu$ in Figure~\ref{fig:cscGX} ($b$). Both ($a$) and ($b$) indicate that ultimate alignment precision is not yet reached. This underlines the importance of including additional external constraints.

\section{Hardware-based Alignment Using Frequency Scanning Interferometry}\label{sec:hardwareBasedAlignment}
\begin{figure*}
\vspace*{-3mm}
\centering
\includegraphics[width=125mm,clip]{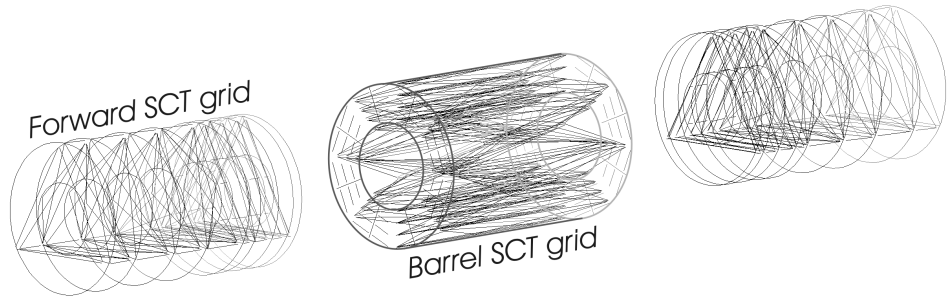}
\vspace*{-4mm}
\caption{\label{fig:fsiGrid}
The geodetic grid of the FSI system (straight lines). The circular lines are for eye-guidance only.
}
\end{figure*}

The SCT has beeen equipped with a Frequency Scanning Interferometry (FSI) optical alignment system~\cite{bib:fsi} which is capable of monitoring micron-scale motions of the detector on a time scale as short as 10\,min. As displayed in Figure~\ref{fig:fsiGrid}, it consists of a geodetic grid of 842 length measurements between nodes, split in barrel and endcaps parts. It is expected that this system will be capable of monitoring low-frequency spatial movements of the SCT with a precision better than 5\,\mum{} in the critical direction~\cite{bib:fsiThesis}.

\begin{figure}
\begin{center}
\begin{picture}(440,110)(0,0)
\put(-30,0){\includegraphics[height=40mm,clip]{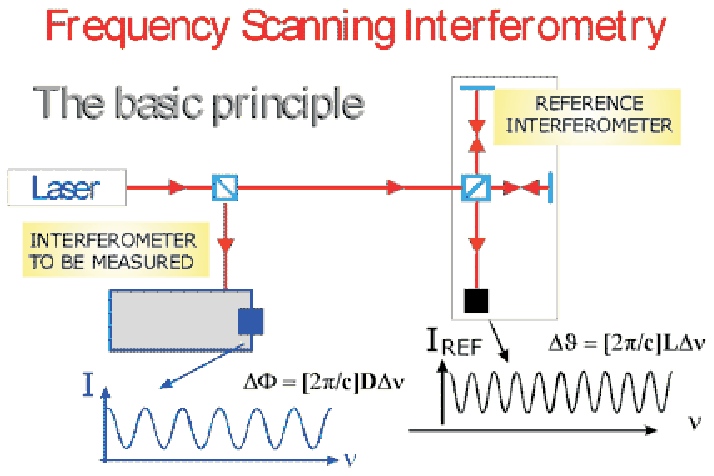}}
\put(160,0){\includegraphics[height=40mm,clip]{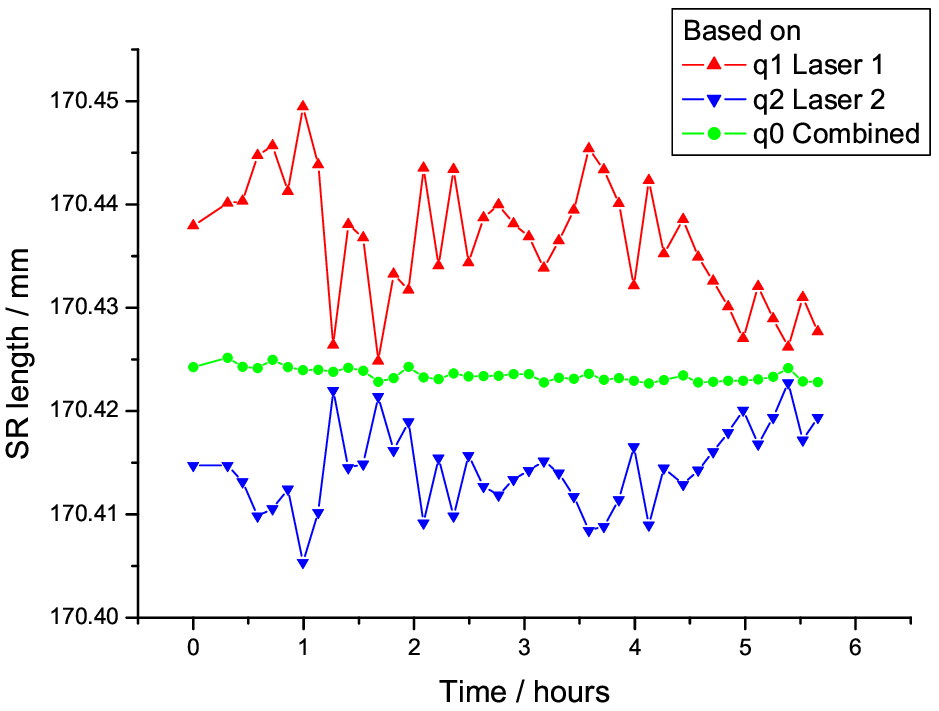}}
\put(330,0){\includegraphics[height=40mm,clip]{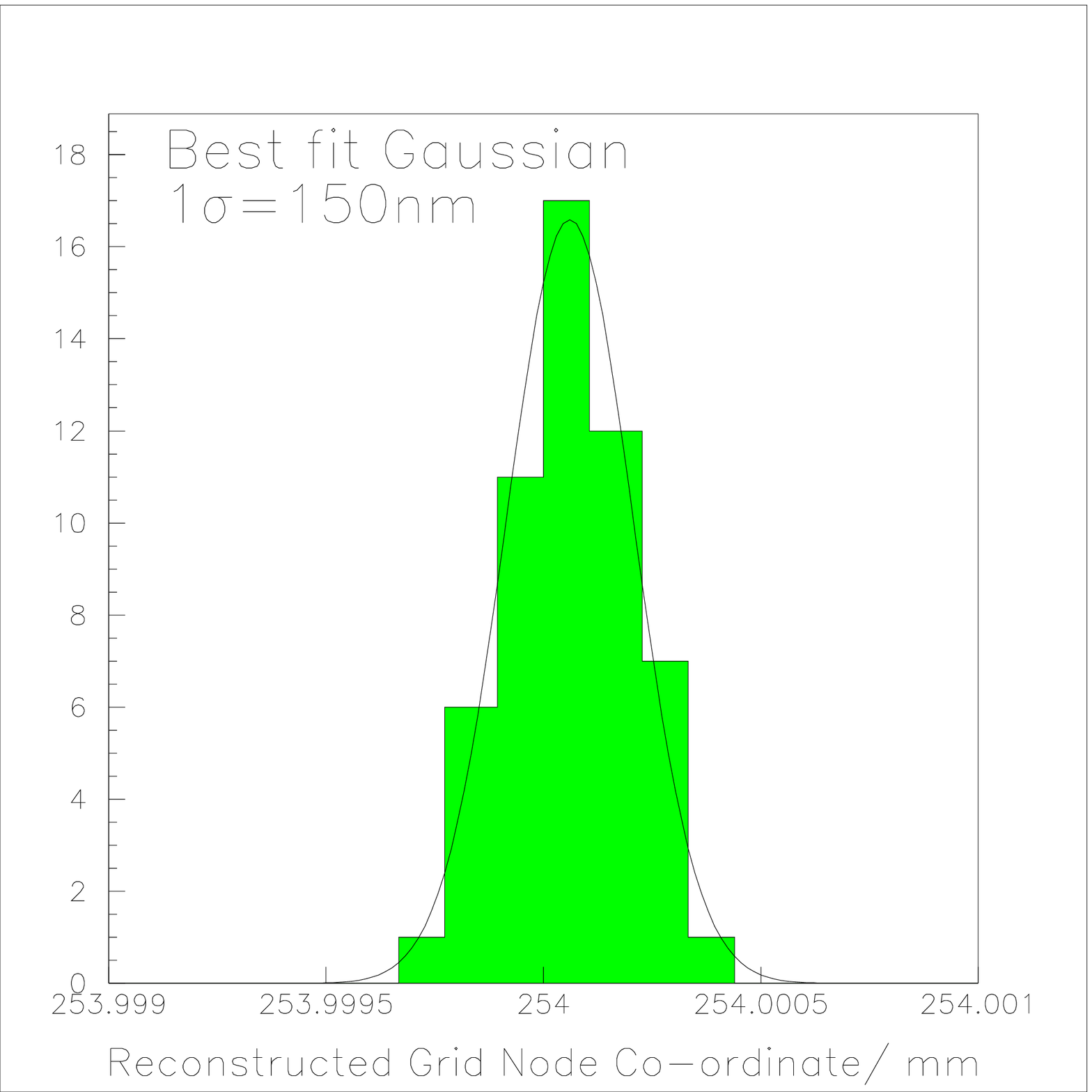}}
\put(-30,22){($a$)}
\put(187,22){($b$)}
\put(350,22){($c$)}
\end{picture}
\vspace*{-3mm}
\caption{\label{fig:fsiDetail}
The basic priciple of the FSI system ($a$), drift cancellation ($b$), and the resolution of the prototype grid ($c$).
}
\vspace*{-0.8cm}
\end{center}
\end{figure}

The principle of the FSI system is explained in Figure~\ref{fig:fsiDetail}~($a$). A tunable laser illuminates multiple fibre-coupled interferometers of the geodetic grid and the reference one, which is located inside an evacuated chamber in a thermally stabilised room. 
As the optical frequency is scanned, a phase shift is introduced in all interferometers, at a rate that is proportional to the length of each interferometer. By comparing the phase shift between the measured grid line and the reference its length can be determined to better than $<$1\,\mum. In practice, the frequency of two lasers is tuned in opposite directions to cancel the effects of drift in the interferometer length, as shown in Figure~\ref{fig:fsiDetail}~($b$). So far, a node precision of 150\,nm has been demonstrated with much smaller prototype grids. 

The installation of the FSI system is almost completed now, and tests are underway. It is expected, that this system will be a vital complement to the track-based alignment algorithms, as it monitors the detector at a much shorter time scale and is sensitive to low-frequency spatial deformation modes.

\vspace{-3mm}
\begin{acknowledgments}
I would like to thank all of the ATLAS Inner Detector alignment community for their help and collaboration, and especially the members of the Oxford group.
\vspace{-3mm}
\end{acknowledgments}


\end{document}